\documentclass[a4paper,french,10pt]{rnti}
\usepackage{amsfonts}
\usepackage{amssymb}
\usepackage{amsmath}
\usepackage{graphicx}
\usepackage{epsfig}

\makeatletter    

\newcommand{\unnumberedcaption}%
	{\@dblarg{\@unnumberedcaption\@captype}}

\newcommand{\@unnumberedcaption}{}
\long\def\@unnumberedcaption#1[#2]#3{\par
  \addcontentsline{\csname ext@#1\endcsname}{#1}{%
    \protect\numberline{}{\ignorespaces #2}%
    }%
  \begingroup
    \@parboxrestore
    \normalsize
    \@makeunnumberedcaption{\ignorespaces #3}\par
  \endgroup}

\newcommand{\@makeunnumberedcaption}[1]{%
  \vskip\abovecaptionskip
  \sbox\@tempboxa{#1}%
  \ifdim \wd\@tempboxa >\hsize
    #1\par
  \else
    \global \@minipagefalse
    \hbox to\hsize{\hfil\box\@tempboxa\hfil}%
  \fi
  \vskip\belowcaptionskip}

\@ifundefined{abovecaptionskip}{%
  \newlength{\abovecaptionskip}%
  \setlength{\abovecaptionskip}{10pt}%
}{}
\@ifundefined{belowcaptionskip}{%
  \newlength{\belowcaptionskip}%
  \setlength{\belowcaptionskip}{0pt}%
}{}

\makeatother    

\newtheorem{defi}{D\'efinition}

\newtheorem{theorem}{Th\'eor\`eme}

\titrecourt{Partitionnement d'un réseau de sociabilité à fort clustering}
\nomcourt{R. Boulet, B. Jouve}
\titre{Partitionnement d'un réseau de sociabilité à fort coefficient de clustering}
\auteur{Romain Boulet,
        Bertrand Jouve
        }
\affiliation{
Institut de Mathématiques de Toulouse,\\
   Université Toulouse II le Mirail\\
    5 Allées Antonio Machado, 31058 Toulouse Cedex 1.\\
          \{boulet,jouve\}@univ-tlse2.fr
          }  
\resume{Afin de comparer l'organisation sociale d'une paysannerie médiévale avant et après la guerre de Cent Ans nous étudions la structure de réseaux sociaux construits à partir d'un corpus de contrats agraires. Faibles diamètres et fort clustering révèlent des graphes en petit monde. Comme beaucoup de grands réseaux d'interaction étudiés ces dernières années ces graphes sont sans échelle typique. Les distributions des degrés de leurs sommets sont bien ajustées par une loi de puissance tronquée par une coupure exponentielle. Ils possèdent en outre un club-huppé, c'est à dire un noyau dense et de faible diamètre regroupant les individus à forts degrés. La forme particulière des éléments propres du laplacien permet d'extraire des communautés qui se répartissent en étoile autour du club huppé. 
}

\summary{
In order to compare social organization of a medieval peasantry before and after the Hundred Years' War we study the sructure of social networks built from a corpus of agrarian contracts. Low diameters and high clusterings show small-world graphs. Like many other networks studied these last years these graphs are scale-free. The distributions of the vertex degrees are fitted by a truncated power law. Moreover they have a rich-club : a dense core with a low diameter consisting of vertices with high degree. The particular shape of the laplacian spectrum allows us to extract communities that are spread along a star whose center is the rich-club. 
}

\Fr
\begin{document}

\section{Introduction}
\footnotetext[0]{Tous les calculs ont été effectués avec le logiciel libre R.}
Les réseaux sociaux sont des systèmes complexes dont certains ont des structures maintenant bien identifiées : graphes de petits mondes et graphes sans échelle typique. Un graphe sans échelle typique est un graphe dont la distribution des degrés n'est pas groupée autour d'une valeur moyenne ; c'est le cas lorsque celle-ci suit une loi de puissance. Les études menées sur le  world wide web, 
des réseaux de courrier électronique 
ou des réseaux P2P 
, le réseau des collaborations scientifiques 
, le réseau des relations sexuelles 
 en sont des exemples \citep{handbook_graphs}.
Les graphes sans échelle typique ont peu de sommets de degrés très élevés et beaucoup de faible degré, ces graphes ont la propriété de présenter des fluctuations locales des degrés d'autant plus importantes que la distribution des degrés est proche d'une loi de puissance.
Si les sommets de forts degrés sont connectés entre eux on parle alors de phénomène de "club huppé" 

Alors que les études ont en général été effectuées sur des réseaux sociaux contemporains nous analysons ici un réseau relatif à la paysannerie médiévale. Nous travaillons sur une base de contrats agraires signés d'une part entre 1240 et 1350 et d'autre part entre 1450 et 1520 dans une petite région du Sud-Ouest de la France. Cette base pour l'instant réduite à environ 700 actes sera amenée à plus de 8000 actes lorsque le travail de saisie et de désambiguïsation sera terminé. Les sommets du graphes sont les paysans et ils sont liés s'ils apparaissent dans un même contrat, nous définissons ainsi deux graphes $G_{av}$ et $G_{ap}$ ; nous avons éclairci la base en enlevant les seigneurs de notre étude. Nous ne possédons pas de données entre 1350 et 1450, intervalle temporel correspondant à la guerre de Cent Ans. 

La notion de communauté varie en fonction du réseau que l'on étudie \citep{palla-2005-435,newman-2006-103}. Nous supposerons dans notre étude que les communautés sont constitués d'individus qui ont à la fois les mêmes liens à l'intérieur de la communauté (clique) et à l'extérieur de la communauté. Nous verrons dans la section \ref{rech_com} que cette définition assez contraignante permet pourtant de révéler une structuration très particulière de notre réseau.
  
Cet article est constitué de deux parties : 
nous allons tout d'abord commencer par vérifier si notre graphe partage les propriétés rencontrées dans les grands réseaux d'interaction (l'effet petit monde, la distribution des degrés et le phénomène de club huppé). Ensuite nous nous attarderons sur la détection et l'organisation des communautés grâce à des méthodes spectrales. Enfin en conclusion, nous ébaucherons une comparaison des graphes avant et après la guerre de Cent Ans.

\section{Les indices des deux réseaux d'interaction}

\subsection{L'effet petit monde}
 
L'effet petit monde regroupe deux propriétés : la première énoncant que la distance entre deux sommets quelconques est faible (ceci est relatif à la connectivité globale) et la deuxième que la connectivité locale est forte. Pour quantifier ces notions nous utiliserons dans le premier cas soit la moyenne des plus courts chemins $\bar{l}$ soit la longueur caractéristique $L$ (médiane des moyennes des plus courts chemins de chaque sommet \citep{small_worlds}) et dans le deuxième cas la moyenne $C_1$ des densités du graphe des voisins de chaque sommet \citep{small_worlds}.
 
Le tableau {\sc Tab}. \ref{petit_monde} résume les résultats obtenus sur les graphes des liens de sociabilités paysans avant et après la guerre de Cent Ans et les met en perspective avec d'autres exemples de réseaux dont les densités d'arêtes sont voisines. Dans le cas de nos deux réseaux signalons qu'ils ont respectivement un diamètre de $5$ et de $6$ et que $90\%$ des paires de sommets sont à une distance inférieure ou égale à $3$.

\begin{center}
\begin{table}[htb]
\begin{center}
\begin{tabular}{|l|c|c|c|c|c|c|}
&$n$ & $|E|$ & $\bar{l}$ & $L$ & densité & $C_1$  \\
\hline
$G_{av}$ & 205 & 1928 &  2.38 & 2.37 & 0.092 & 0.85  \\ 
$G_{ap}$ & 173 & 1044 &  2.52 & 2.47 & 0.070 & 0.85  \\ 
Collaboration entre physiciens $^{(a)}$ & 107&757&2.48&&0.13&0.72\\
Réseau proie/prédateur $^{(b)}$ & 134 & 583  & 2.28 & & 0.065 & 0.21  \\
C.elegans $^{(c)}$ & 282&1974& & 2.65 & 0.05 & 0.28 \\
\hline

\end{tabular}
\end{center}
\caption{Etude de l'effet petit monde sur les graphes $G_{av}$ et $G_{ap}$. $^{(a)}$\cite{small_world_file_sharing}, $^{(b)}$\cite{montoya}, $^{(c)}$\cite{DJW_nature}.}
\label{petit_monde}
\end{table} 
\end{center}

Les coefficients de clustering de nos réseaux sont sensiblement plus forts que ceux rencontrés habituellement.

\subsection{La distribution des degrés}

L'objectif de cette partie est de modéliser la distribution des degrés de nos graphes.

Afin de lisser les fluctuations nous étudions la distribution cumulative des degrés $P_c(k)=\sum_{j=k}^{\infty}P(j)$ où $P(j)$ est la probabilité d'avoir un sommet de degré $j$. Ce choix permet aussi de repérer plus aisément un degré de coupure éventuel au delà duquel la distribution décroit plus vite \citep{internet_structure_evolution}. Si beaucoup de réseaux récemment étudiés montrent une distribution cumulative des degrés qui suit une loi de puissance, la présence d'un degré de coupure est le signe d'un écart à cette loi. L' ajustement de la distribution par une loi de puissance tronquée par une coupure exponentielle (TPL) peut alors permettre de mieux expliquer l'ensemble de la distribution ; citons par exemple \citep{classes_small_world,neuroscience_Achard}. 

\begin{figure}[htb]
 \begin{minipage}[b]{.46\linewidth}
  \centering\epsfig{figure=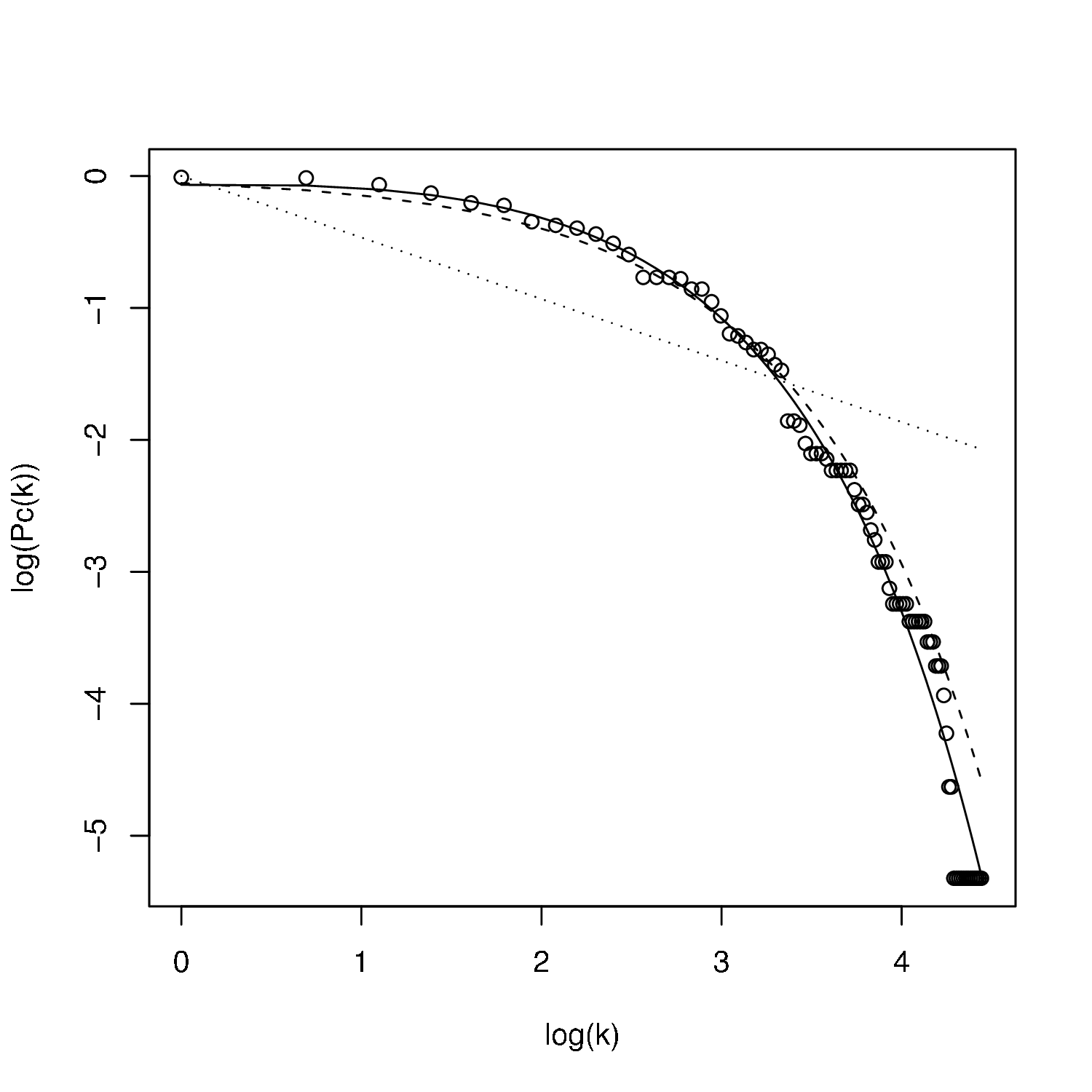,width=\linewidth}
  \unnumberedcaption{(a) Avant la guerre de Cent Ans }
 \end{minipage} \hfill
 \begin{minipage}[b]{.46\linewidth}
  \centering\epsfig{figure=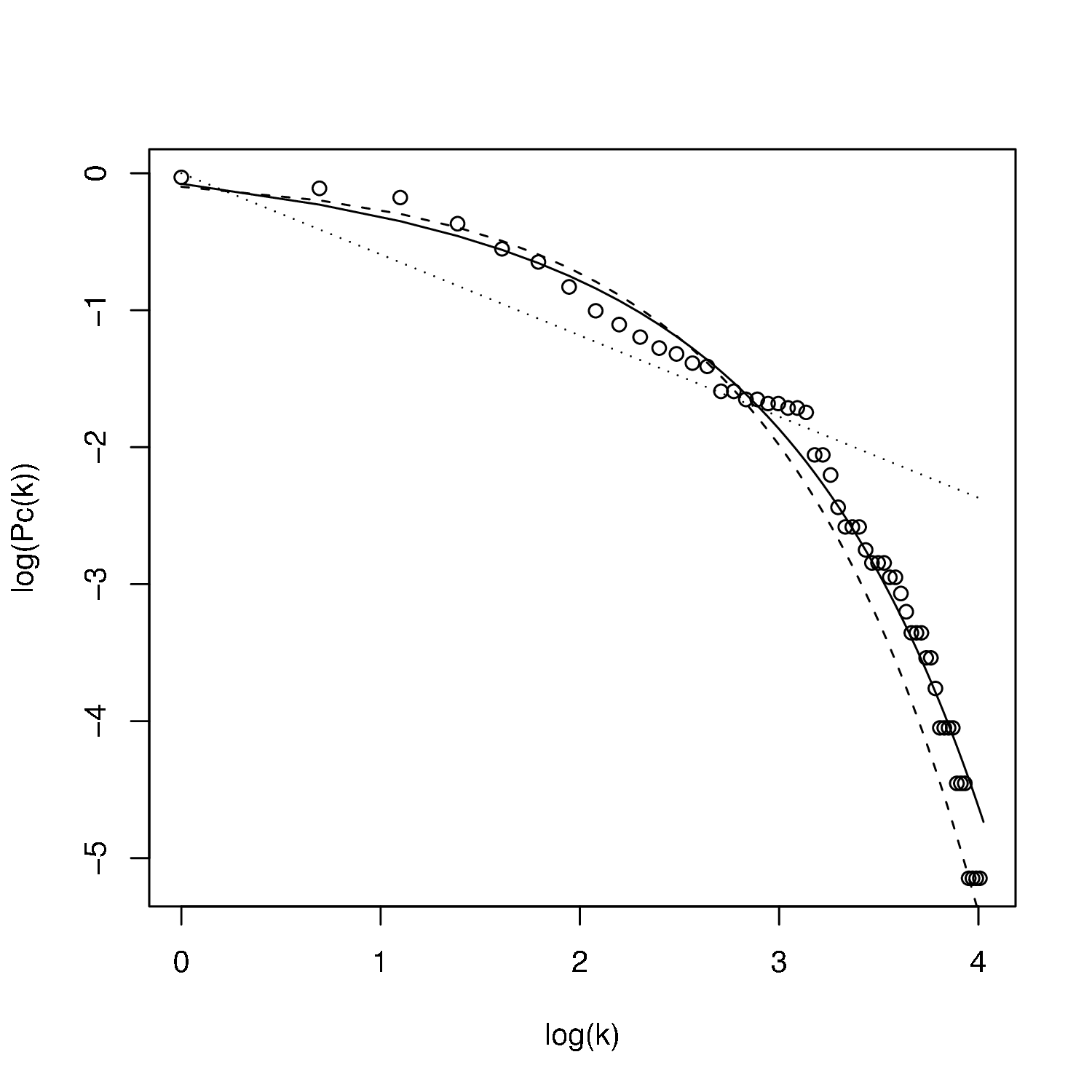,width=\linewidth}
  \unnumberedcaption{(b) Après la guerre de Cent Ans }
 \end{minipage}
 \caption{Distribution cumulative des degrés. En échelle logarithmique sont représentés en abscisse les degrés et en ordonnée la distribution cumulée $P_c$. Les cercles représentent les données, la ligne continue l'approximation par une TPL, la ligne discontinue  par une exponentielle et la ligne en pointillés par une loi de puissance.}
 \label{cumul}
\end{figure}

Nous testons trois lois pour ajuster la distribution : loi de puissance $P_c(k)\sim k^{-\gamma}$, loi exponentielle $P_c(k)\sim e^{-\alpha k}$ et TPL $P_c(k)\sim k^{-\gamma}e^{-\frac{k}{k_c}}$. La figure {\sc Fig}. \ref{cumul} présente les résulats. Dans chacun des cas nous estimons les paramètres au sens des moindres carrés (cf. {\sc Tab}. \ref{tab_coef_err}).  

\begin{center}
\begin{table}[htb]
\begin{center}
\begin{tabular}{|c|c|c|c|c|c|c|c|}
\cline{2-8}
\multicolumn{1}{c}{}&\multicolumn{2}{|c|}{puissance}&\multicolumn{2}{|c|}{exponentielle}&\multicolumn{3}{|c|}{TPL}\\
\cline{2-8}
\multicolumn{1}{c|}{}&erreur&$\gamma$&erreur&$\alpha$&erreur&$\gamma$& $k_c$\\ 
\hline
$G_{av}$ & 22.73 & 0.466 & 0.81 & 0.054 & 0.30 & -0.089&14.91\\
 $G_{ap}$& 8.96 & 0.592 & 1.38 & 0.099 & 1.11  & 0.109& 13.04\\ 
\hline
\end{tabular}
\end{center}
\caption{Coefficients et erreurs quadratiques moyennes ($10^{-3}$) des différents modèles d'ajustement de la distribution cumulative des degrés.}
\label{tab_coef_err}
\end{table}
\end{center}

Le meilleur ajustement de nos données est obtenu pour une TPL. Le graphe $G_{av}$ échappe à une distribution des degrés en loi de puissance. Cette distribution est assez bien ajustée par une loi exponentielle même si on améliore l'erreur quadratique avec une TPL. Pour $G_{ap}$ une loi de faible puissance donne de meilleurs résultats que pour $G_{av}$ mais une TPL reste la meilleure modélisation.

\subsection{L'effet \og club huppé\fg}

Nos deux graphes $G_{av}$ et $G_{ap}$ possèdent un club-huppé \citep{zhou_mondragon_rich_club} c'est-à-dire que les sommets de forts degrés (\og les riches\fg) forment ensemble un sous-graphe dense. 
Ces individus jouant un rôle important dans l'organisation du réseau, les indices de centralité de proximité et de centralité d'intermédiarité \citep{les_resaux_sociaux} nous donnent un critère supplémentaire à celui de la densité et du degré pour le choix des individus du club-huppé. L'indice de centralité de proximité d'un sommet $i$ est l'inverse de la somme des plus courts chemins de $i$ aux autres sommets du graphe. L'indice de centralité d'intermédiarité d'un sommet $i$ est le nombre des plus courts chemins du graphe passant par $i$.
On classe les individus par ordre décroissant des degrés et on retient dans le club huppé ceux dont les indices de centralité sont élevés. Ceci nous conduit à retenir $25$ individus dans $G_{av}$ et $18$ dans $G_{ap}$. 
Le diamètre des clubs huppés est de 2 et leurs densités sont respectivement $0.67$ et $0.81$.

\section{Recherche des communautés}
\label{rech_com}

L'effet petit monde avec un coefficient de clustering élevé associé à une faible densité du graphe nous indique la présence de communautés ; afin de les déceler nous étudions le spectre du laplacien (non normalisé).
 Nous définissons nos communautés ainsi :

\begin{defi}
Une $k$-communauté d'un graphe $G$ est une clique d'ordre $k$ de $G$ (clique non maximale) telle que tous les sommets de cette clique aient les mêmes voisins.
\end{defi}

Nous utiliserons un théorème énoncé par \cite{laplacian_communaute} démontrant que si le laplacien $L$ du graphe a une valeur propre $\lambda$ de multiplicité $k-1$ dont les vecteurs propres associés 
possèdent exactement les mêmes $k$ coordonnées non nulles alors ces $k$ coordonnées correspondent aux sommets d'une $k$-communauté, au sens où nous l'avons définie.

Ces valeurs propres sont nécessairement entières. Considérons en effet $u$ un tel vecteur propre nous avons $Lu=(D-A)u=\lambda u$ et il est par ailleurs facile de voir que $Au=-u$. Ainsi $Du=(\lambda-1)u$ et $\lambda$ est entier.

Nous pouvons énoncer un complément à ce théorème : 

\begin{theorem}
\label{th1}
(i) S'il existe une valeur propre $\lambda$ de $L$ de multiplicité $k-1$ dont les vecteurs propres associés 
possèdent les mêmes $k+1$ coordonnées non nulles  et sont vecteurs propres de la matrice d'adjacence $A$ associé à la valeur propre $-1$ alors il existe deux communautés d'ordre $k_1\geq2$ et $k_2\geq2$ tels que $k_1+k_2=k+1$.

(ii) S'il existe une valeur propre de $L$ de multiplicité $k-1$ dont les vecteurs propres associés 
possèdent les mêmes $k+2$ coordonnées non nulles et sont vecteurs propres de la matrice d'adjacence $A$ associé à la valeur propre $-1$  alors il existe trois communautés d'ordre $k_1$, $k_2$, $k_3$ telles que $k_1+k_2+k_3=k+2$.
\end{theorem}

La démonstration consiste à étudier la matrice binaire $A+I$ qui est de rang 2 dans le cas (i) et de rang 3 dans le cas (ii). On procède par épuisement des cas.

En utilisant le théorème énoncé dans \cite{laplacian_communaute} et le théorème \ref{th1} précédent, nous extrayons pour le graphe $G_{av}$ 28 communautés de taille supérieure ou égale à 3 dont la plus importante est de taille 15 et  pour le graphe $G_{ap}$ $31$ communautés dont la plus grande est de taille 7.

\begin{figure}[htb]
 \begin{minipage}[b]{.46\linewidth}
  \centering\epsfig{figure=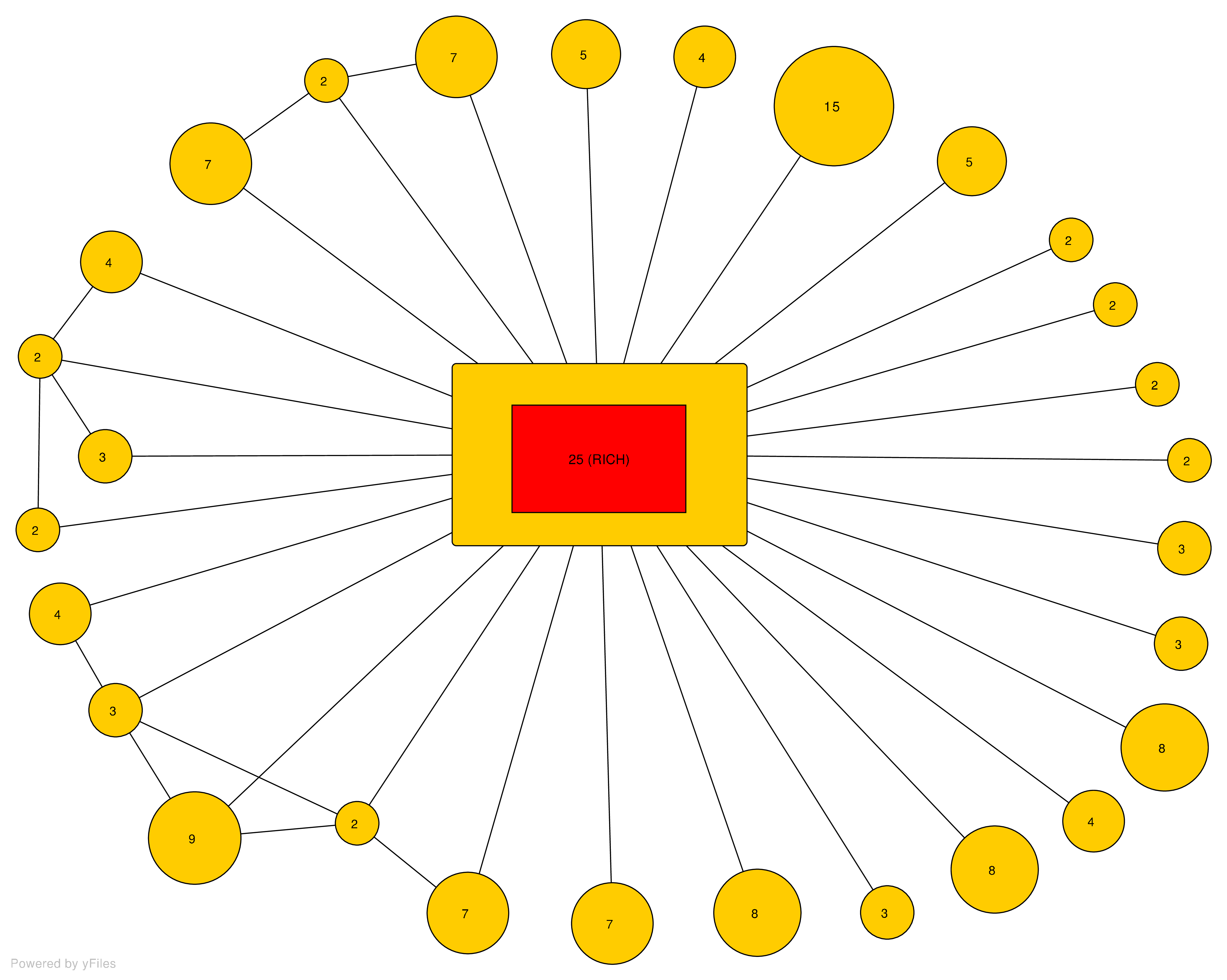,scale=0.14}
 \end{minipage} \hfill
 \begin{minipage}[b]{.46\linewidth}
  \centering\epsfig{figure=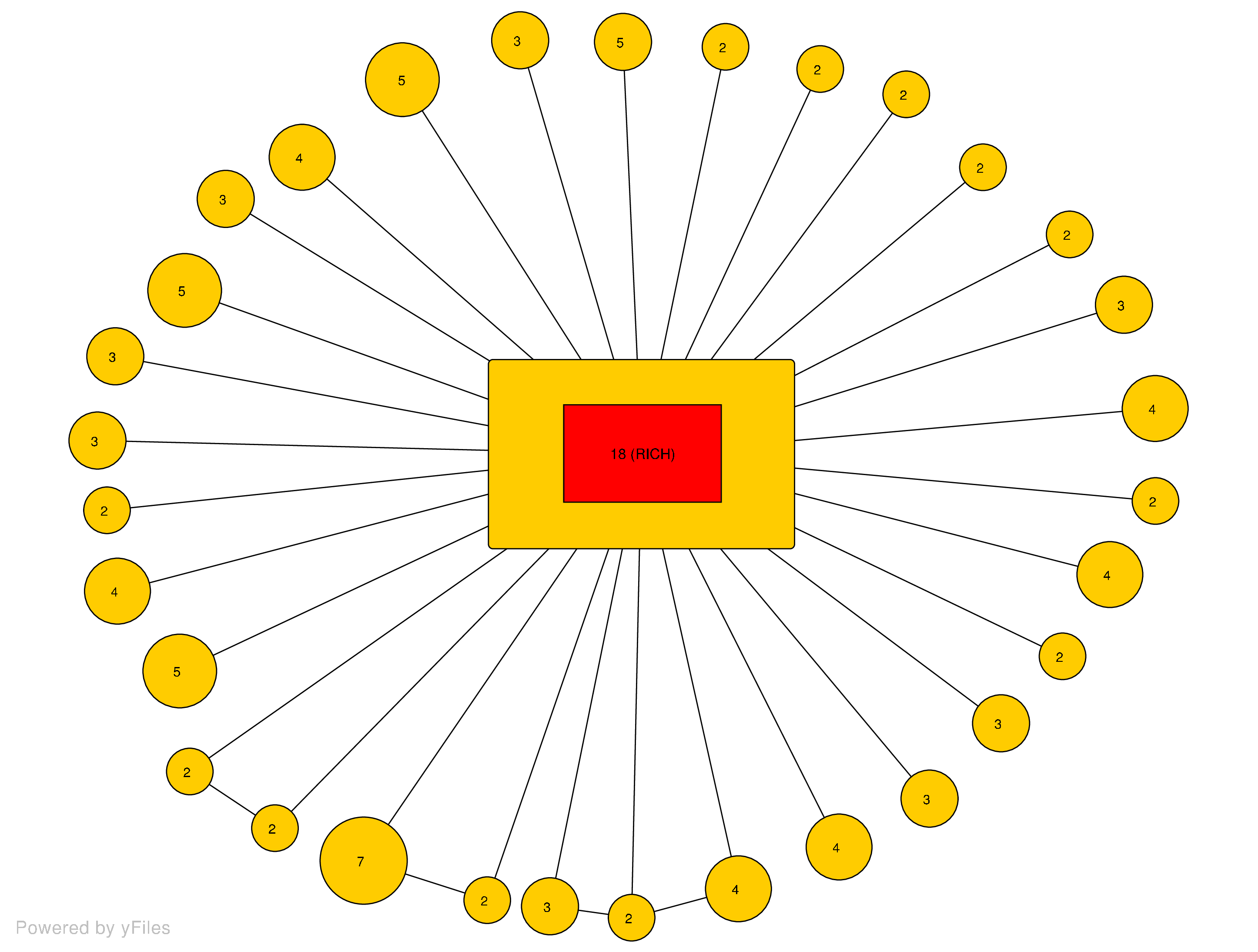,scale=0.19}
 \end{minipage}
 \caption{Graphe des communautés de $G_{av}$ (à gauche) et $G_{ap}$ (à droite). 
 Les disques représentent les $k$-communautés extraites et le rectangle le reste des sommets (dont le club-huppé).
 }
 \label{comm}
\end{figure}

En supprimant la partie du graphe ne contenant aucune  communautés (cette partie contient le club-huppé) nous obtenons un graphe à plusieurs composantes connexes. Les communautés trouvées via l'étude du spectre ne sont donc guère liées entre elles, elles sont préférentiellement liées à la partie que nous avons otée et notamment au club huppé. Nous avons donc une structure inter-communautaire proche de celle d'une étoile. Le centre de cette étoile contient le club-huppé dont on visualise bien à présent le rôle central qu'il joue dans l'organisation du réseau social. Ce partitionnement en club-huppé et communautés permet une bonne visualisation des graphes ({\sc Fig.} \ref{comm})

\section{Conclusion}

Malgré le fait d'avoir enlevé les seigneurs de notre étude, nous constatons dans chacun des deux graphes $G_{av}$ et $G_{ap}$ la présence d'un groupe d'individus (le club huppé) possédant un rôle central. Ces deux graphes apparaissent sous forme d'une étoile de communautés. Dans $G_{av}$, un fort nombre de communautés de petite taille cohabitent avec un nombre significatif de communautés plus importantes, ce qui explique le bon ajustement des distributions des degrés avec une TPL. Concernant $G_{ap}$, la structuration est moins claire : le résidu est sensiblement plus important et la taille des communautés moins variable.

Si certaines de ces communautés correspondent à des zones géographiques comme on peut le voir dans \citep{florent}, d'autres n'ont pour l'instant pas trouvé d'explications.

On remarquera un renouvellement quasi-complet des noms du club-huppé entre $G_{av}$ et $G_{ap}$ : la famille Combelcau très influente avant la guerre disparaît complètement après la guerre laissant la place à la famille Limairac, nouvelle famille qui paraît très influente.

\bibliographystyle{rnti}
\bibliography{./MaBiblio}
 
\end{document}